\documentclass{jps-cp}
\usepackage{txfonts} 

\title{Development of Energy-Resolved Neutron Imaging Detectors at RADEN}

\author{Joseph Don \textsc{Parker}$^{1}$, Masahide \textsc{Harada}$^{2}$, Hirotoshi \textsc{Hayashida}$^{1}$, Kosuke \textsc{Hiroi}$^{2}$,\\ Tetsuya \textsc{Kai}$^{2}$, Yoshihiro \textsc{Matsumoto}$^{1}$, Takeshi \textsc{Nakatani}$^{2}$, Kenichi \textsc{Oikawa}$^{2}$,\\ Mariko \textsc{Segawa}$^{2}$, Takenao \textsc{Shinohara}$^{2}$, Yuhua \textsc{Su}$^{2}$, Atsushi \textsc{Takada}$^{3}$, Taito \textsc{Takemura}$^{3}$,\\ Tomoyuki \textsc{Taniguchi}$^{3}$, Toru \textsc{Tanimori}$^{3}$, and Yoshiaki \textsc{Kiyanagi}$^{4}$}

\inst{$^{1}$Neutron Science and Technology Center, Comprehensive Research Organization for Science and Society (CROSS), Tokai, Ibaraki 319-1106, Japan\\
  $^{2}$J-PARC Center, Japan Atomic Energy Agency, Tokai, Ibaraki 319-1195, Japan\\
  $^{3}$Department of Physics, Kyoto Univeristy, Kyoto 606-8502, Japan\\
  $^{4}$Graduate School of Engineering, Nagoya University, Nagoya, Aichi 464-8603, Japan}

\email{j{\_}parker@cross.or.jp}

\recdate{September 30, 2017}

\abst{
  Energy-resolved neutron imaging at a pulsed source
  utilizes the energy-dependent neutron transmission measured via
  time-of-flight to
  extract quantitative information about the internal microstructure
  of an object.
  At the RADEN instrument at J-PARC in Japan,
  we use cutting-edge detectors
  employing micro-pattern detectors or fast Li-glass scintillators
  and fast, all-digital data acquisition to perform such measurements,
  while continuing their development
  toward better utilization of the intense neutron source.
  In particular, for the Micro-Pixel Chamber
  based Neutron Imaging Detector ($\mu$NID),
  a micro-pattern detector with a 400~$\mu$m pitch and employing $^3$He
  for neutron conversion,
  we have successfully improved the spatial resolution from 200 to
  100~$\mu$m, increased the detection efficiency from
  18 to 26\% for thermal neutrons, and increased the maximum count rate
  from 0.4 to 1~Mcps.
  We are also testing a new readout element
  with a 215~$\mu$m pitch for further improved spatial resolution,
  and a $\mu$NID with boron-based neutron converter for
  increased rate performance.
}

\kword{pulsed neutron imaging, micro-pattern detectors, pixel scintillator detectors}

\begin{document}
\maketitle

\section{Introduction}

The Energy-Resolved Neutron Imaging System, RADEN~\cite{shinohara16},
located at beam line BL22 of the Materials and Life Science
Experimental Facility (MLF) at J-PARC in Japan,
is designed to take full advantage of the
high-intensity, pulsed neutron beam of the MLF to perform not only
conventional radiography/tomography, but also more recently developed
{\em energy-resolved} neutron imaging techniques.
These energy-resolved techniques enable observation of
the macroscopic distribution of microscopic properties within bulk materials
{\em in situ}, including crystallographic structure and internal strain
(Bragg-edge transmission~\cite{sato10}), nuclide-specific density and
temperature distributions (neutron resonance absorption~\cite{sato09}),
and internal/external
magnetic fields (pulsed, polarized neutron imaging~\cite{shinohara11}),
by analysis of the energy-dependent neutron transmission point-by-point
over a sample.
Utilizing the low-divergence, pulsed neutron beam at RADEN,
we combine advanced,
two-dimensional neutron detectors featuring fine time resolution
with the determination of neutron energy by the time-of-flight method
to allow observation of the energy-dependent transmission simultaneously
at all points in a single measurement.
The quantitative nature of these techniques and potentially
short measurement times make energy-resolved neutron imaging
at intense, pulsed neutron sources very
attractive for both scientific and industrial applications.

To carry out such measurements in the high-rate, high-background environment
at a pulsed spallation neutron source such as the J-PARC MLF
requires detectors with sub-$\mu$s time and sub-mm spatial resolutions,
excellent background rejection, and high rate capability.
At RADEN, we use cutting-edge detector systems, which have been
developed in Japan, employing micro-pattern detectors or fast Li-glass
scintillators coupled with high-speed,
FPGA (Field Programmable Gate Array)-based data acquisition systems.
As opposed to conventional CCD camera detectors,
these {\em event-type} detectors measure each individual neutron event
to provide the necessary time resolution and event-by-event background
rejection.
Furthermore, the micro-pattern detectors, by virtue of sub-mm strip pitches,
are able to operate at Mcps (mega-counts-per-second) rates and provide
spatial resolutions on par with conventional CCD camera systems,
while the fast decay time of about 100~ns for Li-glass scintillator
potentially allows high overall count rates on the order of 100~Mcps.

In this paper, we introduce the event-type neutron imaging
detectors in use at RADEN
and discuss our ongoing detector development activities,
including results of tests carried out at RADEN.

\section{Event-Type Detectors at RADEN}

The event-type detector systems currently available at RADEN include
two micro-pattern detectors, the $\mu$NID (Micro-pixel chamber based Neutron
Imaging Detector)~\cite{parker13a,parker13b}
and nGEM (boron-coated Gas Electron Multiplier)~\cite{uno12}
developed at Kyoto University and KEK, respectively,
along with a pixelated Li-glass scintillator detector,
the LiTA12 ($^6$Li Time Analyzer, model 2012)~\cite{satoh15} from KEK.
The main features of these detectors are listed in Table~\ref{tab:dets}.
The micro-pattern detectors have a detection area of $10 \times 10$~cm$^2$
and are based on gaseous time projection chambers
with charge amplification provided by
a micro-pixel chamber ($\mu$PIC)~\cite{ochi01} in the case of the $\mu$NID
and multiple, thin-foil
Gas Electron Multipliers (GEMs)~\cite{gemref} for the nGEM.
The $\mu$NID incorporates $^3$He in the gas mixture
for 26\% efficiency at a neutron energy of $E_n = 25.3$~meV.
The nGEM, on the other hand,
uses a 1.2-$\mu$m thick $^{10}$B coating ($>$98\% purity) deposited on
the aluminum drift cathode and both sides of one GEM foil to achieve
10\% efficiency at $E_n = 25.3$~meV.
The LiTA12 is comprised of a $16 \times 16$ array of $^6$Li-impregnated
glass scintillator pixels (type GS20) matched to a
Hamamatsu H9500 multi-anode photomultiplier with a 3~mm anode pitch
and a total area of $4.9 \times 4.9$~cm$^2$.
The Li-glass scintillator provides a thermal neutron efficiency of
more than 48\% per pixel at $E_n = 25.3$~meV,
with an overall efficiency of 23\% when
including dead space between the pixels.
All detectors feature fast, all-digital FPGA-based
data acquisition with data transfer over Gigabit Ethernet
to provide for the necessary time resolution and
high-rate operation required at the intense
pulsed neutron source of the MLF.

In preliminary testing reported in Refs.~\cite{parker15} and \cite{parker16},
as well as in subsequent testing,
the expected spatial resolution of each detector was confirmed at RADEN.
The $\mu$NID and nGEM
were evaluated using a Gd test target designed at RADEN~\cite{segawa17},
while that of the LiTA12 was confirmed using a simple shape
made from Cd plate.
The rate performance of each system was also studied using adjustable
B$_4$C slits to vary the incident neutron intensity.
To characterize the rate performance, we determined two quantities:
{\em peak count-rate capacity}, which indicates the absolute maximum
instantaneous neutron rate measured over the whole detector, 
and {\em effective peak count rate}, which indicates the instantaneous peak
rate achievable over the whole detector
with good linearity in count rate versus incident intensity
(where {\em good} is defined here as less than 2\% event loss).
Both of these rates are what are referred to as
{\em global instantaneous peak rates}~\cite{stefanescu}.
With the strongly-peaked neutron time-of-flight spectrum at the MLF,
it is these global instantaneous rates which limit the
performance of the detector systems at RADEN.
The results of these studies
are listed in Table~\ref{tab:dets}.

\begin{table}[tbh]
\caption{Features of event-type detectors available at RADEN are listed
  below.
  The values for the spatial resolution, peak count-rate capacity, and
  effective peak count rate were confirmed at RADEN.
  (The terms {\em peak count-rate capacity} and {\em effective peak count
    rate} are defined in the text.)
}
\label{tab:dets}
\center
\begin{tabular}{lccc}
\hline
Detector & $\mu$NID & nGEM & LiTA12 \\
\hline
Type & Micro-Pattern & Micro-Pattern & Pixelated Scintillator \\
Neutron converter & $^3$He & $^{10}$B & $^6$Li \\
Area & $10 \times 10$~cm$^2$ & $10 \times 10$~cm$^2$ & $4.9 \times 4.9$~cm$^2$ \\
Time resolution & 0.25~$\mu$s & 15~ns & 40~ns \\
Spatial resolution & 0.1~mm & 1~mm & 3~mm \\
Efficiency (at $E_n = 25.3$~meV) & 26\% & 10\% & 23\% \\
Peak count-rate capacity & 8~Mcps & 4.6~Mcps & 8~Mcps \\
Effective peak count rate & 1~Mcps & 180~kcps & 6~Mcps \\
\hline
\end{tabular}
\end{table}

We are also actively developing the LiTA12 and $\mu$NID at RADEN
in order to optimize their characteristics, including spatial
resolution, rate performance, and detection efficiency, for energy-resolved
neutron imaging.
(While the nGEM is extensively used at RADEN, it is not currently being
developed by our group.)
Specifically, the LiTA12 is being optimized for neutron resonance
absorption measurements by replacing the scintillator pixels with
a single, flat scintillator with increased thickness in order to
increase detection efficiency for epithermal neutrons
(i.e., those in the resonance energy region
above $E_n \simeq 1$~eV)~\cite{kai17}.
The single scintillator also allows calculation of the centroid of
multiple anodes
for an improvement in spatial resolution to less than
1~mm~\cite{kai17,segawa17}.
Additionally, for the $\mu$NID, we have upgraded the data acquisition hardware
and optimized the gas mixture for improved spatial resolution
and rate performance~\cite{parker16},
with further development underway.
For the remainder of this paper, we will discuss ongoing
development of the $\mu$NID in detail.

\section{Development of the $\mu$NID at RADEN}

\begin{figure}[tbh]
\centering
\includegraphics[width=145mm,clip]{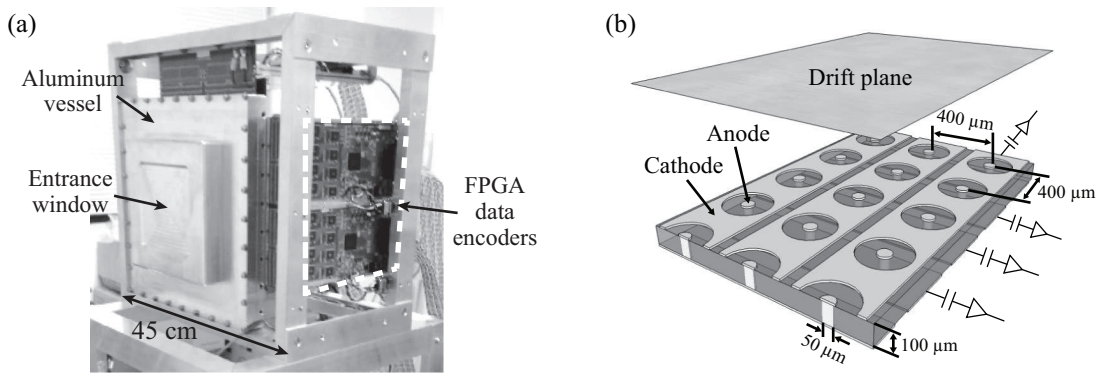}
\caption{A photograph of a $\mu$NID system is shown in (a) with the
  aluminum pressure vessel, entrance window, and FPGA-based data encoder
  modules indicated.
  An illustration of the time-projection chamber showing the drift
  plane and structure of the $\mu$PIC readout is shown in (b)
  (drift plane--$\mu$PIC separation not to scale). 
\label{fig:unid}}
\end{figure}

The $\mu$NID, shown in Fig.~\ref{fig:unid}(a),
uses a time projection chamber with a drift length of 2.5~cm
and a 10~cm $\times$ 10~cm readout plane consisting of a
micro-pixel chamber ($\mu$PIC), coupled to a modular,
FPGA-based data acquisition system~\cite{mizumoto15}.
The $\mu$PIC is a micro-pattern detector with a 400~$\mu$m pitch,
two-dimensional strip readout,
which through its unique microstructure, illustrated in Fig.~\ref{fig:unid}(b),
achieves both charge amplification and analog strip readout.
To facilitate neutron detection, a CF$_4$--iC$_4$H$_{10}$--$^3$He gas
mixture (mixing ratio 45:5:50) at 2~atm total pressure is used,
providing a detection efficiency of 26\% at $E_n = 25.3$~meV
(on par with conventional CCD camera systems).
In this gas mixture, the tracks of the reaction products are less than
5~mm.
A drift field of 1,600~V/cm is used (50~$\mu$m/ns drift velocity, 
0.5~$\mu$s maximum drift time),
and the $\mu$PIC readout is operated at an anode voltage around 650~V
for a gain factor of 100 to 150.
Following a neutron-$^3$He interaction, the three-dimensional track and
energy deposition (estimated via time-over-threshold)
of the resultant
proton-triton pair are recorded in the FPGA-based data encoder modules
and sent to PC via Gigabit Ethernet.
This detailed tracking information allows the $\mu$NID to achieve
a fine spatial resolution of 0.1~mm and a low
gamma sensitivity of less than 10$^{-12}$.
The $\mu$NID also features a time resolution of 0.25~$\mu$s,
a peak count-rate capacity of 8~Mcps,
and an effective peak count rate of 1~Mcps.

In our initial development described in Ref.~\cite{parker16},
we performed an upgrade of the data acquisition hardware
and replaced the original Ar--C$_2$H$_6$--$^3$He (63:7:30 at 2~atm)
gas mixture~\cite{parker13a} with the current CF$_4$-based mixture.
For the hardware,
the data output port of the FPGA-based data encoder modules was upgraded
from 100BASE-T to Gigabit Ethernet, improving the throughput of the
data acquisition hardware by roughly a factor of nine.
The CF$_4$-based gas mixture provided a more than two times faster
drift velocity for shorter charge evacuation times,
nearly two times the stopping power for reduced event sizes,
and a more than three-fold reduction in the electron diffusion for
improved event localization as compared to the Ar-based mixture.
Furthermore, the higher stopping power of CF$_4$ allowed us to increase the
$^3$He fraction while maintaining smaller event sizes.
Taken together, these detector improvements provided an increase in the
peak count-rate capacity from 0.6 to 8~Mcps
and an increase in the detection efficiency from 18\% to the current 26\%.
The updated encoder modules and new gas mixture have been thoroughly tested
at RADEN and are now part of the standard setup for our $\mu$NID system.
Ongoing development efforts to improve the spatial resolution
and rate performance are described below,
including optimization of data analysis algorithms,
development of a new $\mu$PIC readout plane with reduced pitch,
and testing of a $\mu$NID with $^{10}$B-based neutron converter.

\subsection{Optimization of data analysis algorithms}

The digital data produced by the $\mu$NID consists of a stream of hits
comprised of strip number, hit time, and a flag indicating whether
the analog signal was rising or falling when it crossed the
discriminator threshold.
It is the job of the offline data analysis to match rising and falling hits
and calculate the time-over-threshold,
group individual hits into neutron events (referred to as {\em clustering}),
and determine the precise neutron interaction point
(referred to as {\em position reconstruction}).
By optimizing the clustering and position reconstruction algorithms,
we have recently been able to improve the effective peak count rate and
maximize the spatial resolution of the $\mu$NID.

\subsubsection{Clustering algorithm}

\begin{figure}[tbh]
\centering
\includegraphics[width=140mm,clip]{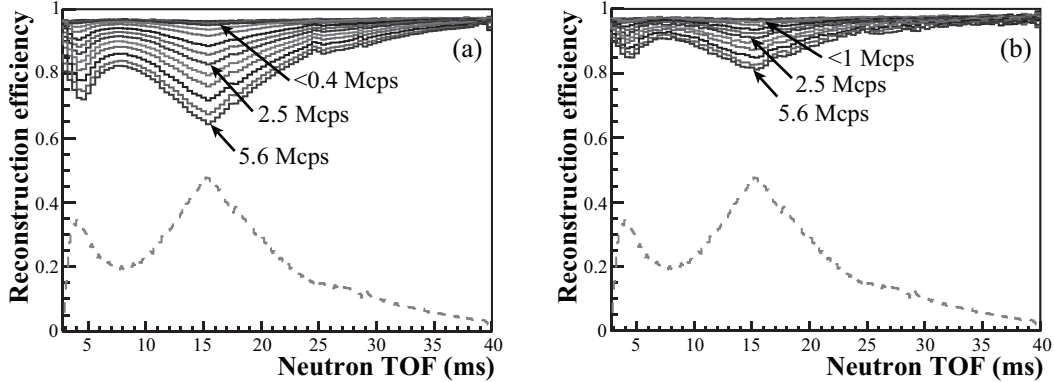}
  \caption{Neutron reconstruction efficiency for the $\mu$NID versus
    neutron time-of-flight is shown for a range of incident neutron
    intensities for:
    a) the original single-linkage clustering algorithm, and
    b) the improved density-based clustering algorithm with explicit
    event pile-up resolution described in the text.
    For reference, a typical neutron TOF spectrum
    is shown as the dashed line in each plot (with arbitrary scale).
\label{fig:neff}}
\end{figure}

After the hardware upgrade of Ref.~\cite{parker16},
the $\mu$NID achieved a peak count-rate capacity of 8~Mcps with
good linearity up to this maximum when considering only raw hits.
The lower effective peak count rate arises mostly from the
clustering of the offline analysis.
The original clustering algorithm was based on a simple, single-linkage
clustering with hits grouped solely by the distance between them
(i.e., all hits whose inter-hit separation was within a specified cut-off
were considered to come from the same event).
While this simple algorithm worked well at low neutron rates,
event pile-up was seen to become significant at global peak rates
above 400~kcps.
This is illustrated in Fig.~\ref{fig:neff}(a),
where the {\em neutron reconstruction efficiency},
defined as the ratio of reconstructed neutron events
to the expected number of neutron events (as derived from the number of
raw hits), is plotted as a function of neutron time-of-flight (TOF)
for global peak rates up to 5.6~Mcps.
The clear dip at the peak of the TOF distribution,
shown by the dashed line in Fig.~\ref{fig:neff}(a), indicates
event loss due to pile-up, which increases with the peak rate.
The observed event loss is about 2\% at a global peak rate of 400~kcps.

To address this poor performance,
we are now developing a new algorithm
employing density-based clustering (based on DBSCAN~\cite{dbscan}),
followed by explicit event pile-up resolution.
For our initial study, clusters that overlap in time were grouped,
and a mismatch in the number of clusters on each of the perpendicular
strip planes was taken as a signal of event pile-up.
Then, in the case that the number of clusters differs by one,
the largest cluster was assumed to be a pile-up event and was allowed
to pair with two clusters from the opposite strip orientation.
Even with this simple method,
the improvement in the neutron reconstruction efficiency
is clearly visible in Fig.~\ref{fig:neff}(b),
where event loss was reduced to 2\% at 1~Mcps global peak rate.
The neutron reconstruction efficiency is expected to improve further
as we increase the sophistication of the event pile-up resolution
algorithm.
We will also study the effect of the new clustering algorithm on
the spatial resolution.

\subsubsection{Position reconstruction algorithm}

\begin{figure}[tbh]
\centering
\includegraphics[width=120mm,clip]{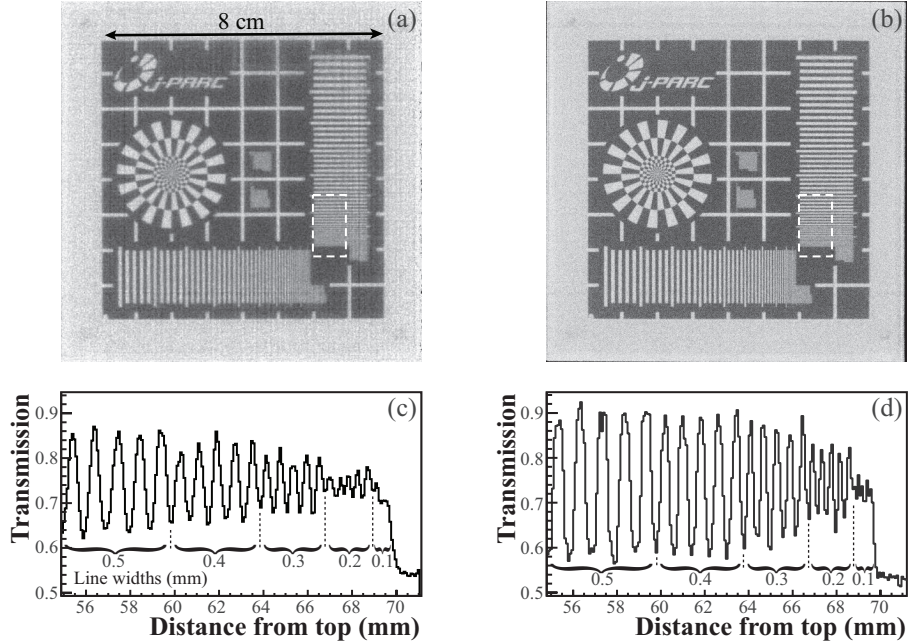}
  \caption{Images of a gadolinium test target taken with the $\mu$NID
    are shown as reconstructed from the same data using:
    a) the original template selection algorithm, and
    b) the improved template selection algorithm described in the text.
    The image area is $10 \times 10$~cm$^2$ with a bin size of
    $40 \times 40$~$\mu$m$^2$ for each.
    Projections of the line pairs within the dashed
    boxes in (a) and (b) are shown in (c) and (d), respectively.
\label{fig:sres}}
\end{figure}

In the offline analysis,
neutron position is determined event-by-event
via a fit to the time-over-threshold (TOT) distributions
for each strip orientation, as described in Ref.~\cite{parker13b}.
The fits are carried out using {\em template} distributions
generated with a GEANT4~\cite{geant4_1,geant4_2} simulation
of the $\mu$NID system,
where the templates are indexed by the proton-triton track unit vector.
This fitting procedure allows the clean separation of the proton and
triton (with $<$5\% observable contamination from misidentified events),
facilitating the fine spatial resolution achieved by this detector.
In the original template selection algorithm, the unit vector was determined
from the fully reconstructed three-dimensional track, requiring
input of three adjustable parameters (i.e., {\em x}, {\em y},
and {\em z} offsets) and several calculation steps.
We have recently developed a simplified template selection algorithm that
uses the track projections, which are measured directly, and one
adjustable parameter, namely, the average proton-triton track length.
Images of the Gd test target taken at RADEN
are shown in Fig.~\ref{fig:sres}(a) and \ref{fig:sres}(b),
reconstructed from the same data using the original and simplified
template selection algorithms, respectively.
Also, projections of the line pairs within the dashed regions in
Figs.~\ref{fig:sres}(a) and \ref{fig:sres}(b) are shown in
Figs.~\ref{fig:sres}(c) and \ref{fig:sres}(d), respectively.
From Figs.~\ref{fig:sres}(c) and \ref{fig:sres}(d),
the spatial resolution was evaluated as 200 and 100~$\mu$m, respectively,
at a Modulation Transfer Function (MTF) value of 10\%.
These results show clearly that the simplified template selection
algorithm provides both improved spatial resolution and image uniformity,
indicating better matching of templates to the measured TOT distributions.

\subsection{$\mu$PIC with reduced strip pitch}

To provide a significant improvement in the spatial resolution,
we are working with the manufacturer of the $\mu$PIC,
DaiNippon Printing Co., Ltd., to develop a new $\mu$PIC readout element
with reduced strip pitch.
(Simulations indicate that the spatial resolution should scale roughly
with the strip pitch.)
The standard $\mu$PIC described above is manufactured using
conventional printed circuit board techniques, which, due to poor
tolerances, are not well suited to producing very fine structures below
several 10s $\mu$m.
By changing to a MEMS (Micro-Electro-Mechanical Systems)-based process,
however, smaller structures (as small as 10~$\mu$m) can be created with very
good uniformity.
Using MEMS manufacturing techniques,
a new $\mu$PIC readout element,
referred to as a TSV (Through-Silicon-Via) $\mu$PIC,
with a 215~$\mu$m pitch,
or nearly half that of the standard, 400-$\mu$m pitch $\mu$PIC,
has been successfully produced.
For our initial study, a 215 $\mu$m pitch test piece,
comprised of $64 \times 64$ strips
for a detection area of $1.4 \times 1.4$~cm$^2$, was manufactured
and testing was carried out at RADEN.
In preliminary testing described in Ref.~\cite{parker16},
the TSV $\mu$PIC test piece provided sufficient gain for neutron detection
(at about 200),
but showed poor gain stability under sustained neutron irradiation.
This observed gain instability was thought to arise from charge build-up within
the silicon substrate, which in the MEMS process is used in place of the
insulating polyimide substrate of the standard $\mu$PIC.

\begin{figure}[tbh]
\centering
\includegraphics[width=127mm,clip]{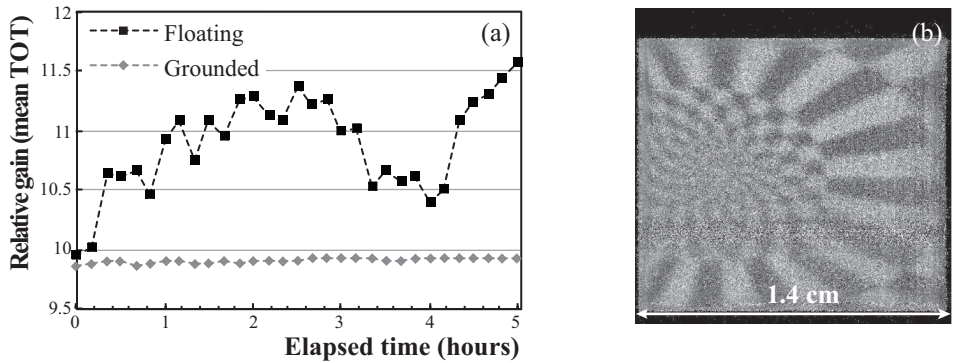}
  \caption{Results of TSV $\mu$PIC tests are shown for:
    a) gain stability with and without substrate grounding, and
    b) imaging of gadolinium test target (Siemens star).
    The image area is 1.4~cm $\times$ 1.4~cm
    (21.5~$\mu$m $\times$ 21.5~$\mu$m bin size).
    The dark areas at top and bottom and distortion in the lower
    third are due to damaged strips.
\label{fig:mems}}
\end{figure}

Based on the above assumption, we investigated the
effect of electrically grounding the silicon substrate,
which would be expected to allow the evacuation of any charge build-up,
in a subsequent test of the TSV $\mu$PIC.
Figure~\ref{fig:mems}(a) shows the relative gain measured over a 5-hour
period of
constant neutron irradiation with and without grounding the substrate,
where the gain is represented by the peak TOT value
averaged over all channels.
These results show that by grounding the substrate,
the gain stability of the TSV $\mu$PIC can be significantly improved.
With the gain stabilized,
we were then able to take the first test
image with the fine-pitch $\mu$PIC as shown in Fig.~\ref{fig:mems}(b).
The statistics are low, but the Siemens star of our Gd test target is
clearly visible.
We are now preparing a new 215~$\mu$m pitch test piece with $256 \times 256$
strips and an area of 5.5~cm $\times$ 5.5~cm, and we will continue
to study the gain stability and spatial resolution with this
large-area TSV $\mu$PIC from this spring at RADEN.

\subsection{$\mu$NID with boron-based neutron converter}

To provide a significant improvement in the peak count-rate capacity,
we are also developing a $\mu$NID with a $^{10}$B-based neutron converter.
Use of a boron-based converter is expected to provide a three-fold
increase in the peak count-rate capacity of the system (to over 20~Mcps)
due to the fact that the
alpha particle released in the neutron-$^{10}$B reaction travels a much
shorter distance in the gas of the detector as compared to the lighter
proton and triton in the $^3$He case,
thereby creating fewer hits per event and
allowing more events to be transmitted over the same system bandwidth.
This small event size (of only 2 or 3 hit strips per readout direction),
however, comes with a trade-off in spatial resolution
as the limited information renders detailed reconstruction algorithms, such
as the template-fitting method above and the $\mu$TPC method of
Ref.~\cite{pfeiffer}, less effective.
The change from $^3$He gas to a $^{10}$B-based converter also reduces
the long-term maintenance costs of the detector.

\begin{figure}[tbh]
\centering
\includegraphics[width=125mm,clip]{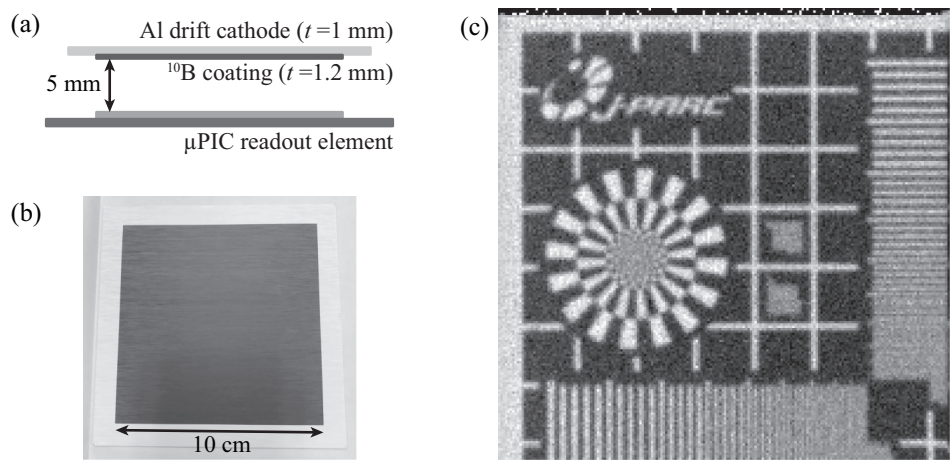}
\caption{
  Shown here are:
  a) a simple diagram showing the basic structure of the $\mu$NID with
  boron converter,
  b) a photograph of the 1.2~$\mu$m $^{10}$B layer (dark rectangular area)
  deposited on one side of the aluminum drift cathode, and
  c) an image of the gadolinium test target taken with the $\mu$NID
  with boron converter at RADEN.
  The image area in (c) is 7.7~cm $\times$ 7.7~cm with a bin size of
  400~$\mu$m $\times$ 400~$\mu$m.
\label{fig:boron}}
\end{figure}

As a proof-of-principle demonstration, we installed an aluminum
drift cathode with a 1.2-$\mu$m thick $^{10}$B coating
($>$98\% purity) into one of our $\mu$NID
systems, as shown in Figs.~\ref{fig:boron}(a) and \ref{fig:boron}(b),
and filled the vessel with a CF$_4$-iC$_4$H$_{10}$ (90:10) gas mixture
at 1.6~atm.
The CF$_4$-based gas mixture was chosen for its high stopping power to
keep the alpha tracks short.
Figure~\ref{fig:boron}(c) is an image of the Gd test target produced
at RADEN,
showing a spatial resolution of around 500~$\mu$m, or slightly larger than
the pitch of the $\mu$PIC strip readout.
We also observed a 2.8 times reduction in event size compared to
the $^3$He case,
confirming, in principle, an expected increase in peak count-rate capacity
of up to 22~Mcps.
Due to the low efficiency of only 3 to 5\% at $E_n = 25.3$~meV for
the present
converter and a limited neutron beam power of 150 kW at the time of the
measurement, however,
we were unable to directly measure the peak count-rate capacity.
We are now considering new converter designs
for increased efficiency, and we will measure the peak count-rate capacity
in a future test at the MLF.

\section{Conclusion}

At the RADEN instrument of the J-PARC MLF, we use advanced event-type
neutron imaging detectors, including the $\mu$NID and nGEM micro-pattern
detectors and the LiTA12 scintillator pixel detector.
The performance of these detectors has been verified at RADEN,
and they have been used by both the RADEN instrument group and general
users to carry out energy-resolved neutron imaging measurements since 2015.
In order to fully utilize the intense pulsed neutron beam of the MLF
and better meet the needs of users, we continue to develop
these detectors for improved spatial resolution, higher efficiency,
and better rate performance.
Specifically, through the ongoing development of the $\mu$NID system described
here, we have improved the spatial resolution from 200
to 100 $\mu$m at 10\% MTF, increased the efficiency from 18 to 26\%
at $E_n = 25.3$~meV, and increased the effective peak count rate from
0.4 to 1~Mcps, with further improvement expected
with optimization of the offline event analysis.
Furthermore, a new 215 $\mu$m pitch $\mu$PIC
is expected (from simulation) to provide nearly double the spatial resolution,
while a $\mu$NID with boron converter should provide a factor of
three increase in peak count-rate capacity for more than 20~Mcps
total throughput.

\section*{Acknowledgment}

The development of
the 215~$\mu$m pitch TSV $\mu$PIC readout element
and the $\mu$NID with boron converter
was partially supported by JST ERATO Grant No. JPMJER1403, Japan.
Testing at RADEN was carried out under
MLF Instrument Group Use Proposal No. 2017I0022,
MLF General Use Proposal No. 2016B0161,
and CROSS Development Use Proposal No. 2017C0004.

\end{document}